\documentclass[aps,prb,twocolumn,superscriptaddress,a4paper]{revtex4-1}
\usepackage[dvipdfmx]{graphicx}

\usepackage{amsmath,amsthm,amssymb}
\usepackage{amsmath,bm}
\usepackage{color}
\usepackage{ifthen}

\usepackage{multirow,bigdelim}

\newcount \commentout
\newcommand{\1}{\mbox{1}\hspace{-0.25em}\mbox{l}}

\newcommand{\sgn}{\mathrm{sgn}}

\newlength{\figwidth}
\newlength{\figlarge}
\begin{document}
\title{
Correlation effects on non-Hermitian point-gap topology in zero dimension: reduction of topological classification
}

\author{Tsuneya Yoshida}
\author{Yasuhiro Hatsugai}
\affiliation{Department of Physics, University of Tsukuba, Ibaraki 305-8571, Japan}
\date{\today}
\begin{abstract}
We analyze a zero-dimensional correlated system with special emphasis on the non-Hermitian point-gap topology protected by chiral symmetry.
Our analysis elucidates that correlations destroy an exceptional point on a topological transition point which separates two topological phases in the non-interacting case; one of them is characterized by the zero-th Chern number $N_{0\mathrm{Ch}}=0$, and the other is characterized by $N_{0\mathrm{Ch}}=2$.
This fact implies that correlations allow to continuously connect the two distinct topological phases in the non-interacting case without closing the point-gap, which is analogous to the reduction of topological classifications by correlations in Hermitian systems.
Furthermore, we also discover a Mott exceptional point, an exceptional point where only spin degrees of freedom are involved.
\end{abstract}
\maketitle


\section{
Introduction
}
\label{sec: intro}

Topological properties of non-Hermitian Bloch Hamiltonians have been extensively addressed as one of recent hot topics in condensed matter physics~\cite{HShen2017_non-Hermi,Gong_class_PRX18,KKawabata_TopoUni_NatComm19,Kawabata_gapped_PRX19,Zhou_gapped_class_PRB19,Bergholtz_Review19,Yoshida_nHReview_PTEP20,Ashida_nHReview_arXiv20}.
For non-Hermitian systems, a variety of novel phenomena have been reported~\cite{Hatano_PRL96,CMBender_PRL98,Fukui_nH_PRB98,Hu_nH_PRB11,Esaki_nH_PRB11,Alvarez_nHSkin_PRB18,SYao_nHSkin-1D_PRL18,SYao_nHSkin-2D_PRL18,KFlore_nHSkin_PRL18,Lee_Skin19,EElizabet_PRBnHSkinHOTI_PRB19,Borgnia_ptGapPRL2020,Yokomizo_BBC_PRL19,Yokomizo_BBC_PRL19}.
In particular, the point-gap topolgy induces topological phenomena which do not have Hermitian counter parts.
For instance, the point-gap topology induces exceptional points (EPs) on which both the real and imaginary parts of the energy eigenvalues touch due to the violation of diagonalizability~\cite{HShen2017_non-Hermi,Budich_SPERs_PRB19,Yoshida_SPERs_PRB19,Okugawa_SPERs_PRB19,Zhou_SPERs_Optica19,Kimura_SPERs_PRB19,Kawabata_gapless_PRL19,Yang_Descri_PRL21,Delplace_Resul_arXiv21}.
In addition, the point-gap topology induces skin effects which are described by the novel bulk-boundary correspondence~\cite{Zhang_BECskin19,Okuma_BECskin19}; their ubiquity also became clear recently~\cite{Hofmann_ExpRecipSkin_19,Xiao_nHSkin_Exp_NatPhys19,Yoshida_MSkinPRR20,Okugawa_HOSkin_arXiv2020,Kawabata_HOSkin_arXiv2020,Fu_HOSkin_arXiv2020,Kawabata_TQFTSkin_arXiv20}.
Platforms of the above non-Hermitian topology extends to a wide range of systems, such as open quantum systems~\cite{TELeePRL16_Half_quantized,YXuPRL17_exceptional_ring,Diehl_DissCher_NatPhys11,Bardyn_DissCher_NJP2013,Lieu_Liouclass_PRL20}, photonic crystals~\cite{Guo_nHExp_PRL09,Ruter_nHExp_NatPhys10,Regensburger_nHExp_Nat12,Zhen_AcciEP_Nat15,Hassan_EP_PRL17,Takata_pSSH_PRL18,Zhou_ObEP_Science18,Zhou_FermiArcPH_Science18,Ozawa_TopoPhoto_RMP19}, mechanical metamaterials~\cite{Yoshida_SPERs_mech19,Ghatak_Mech_nHskin_PNAS20,Scheibner_nHmech_PRL20}, quasi-particles in equibium systems~\cite{VKozii_nH_arXiv17,Zyuzin_nHEP_PRB18,Yoshida_EP_DMFT_PRB18,HShen2018quantum_osci,Papaji_nHEP_PRB19,Matsushita_ER_PRB19,Matsushita_nHResp_arXiv20}, and so on.

Along with the above crucial progress in the non-interacting case, it turned out that correlation effects enrich topological phenomena as is the case of Hermitian sysetms~\cite{Yoshida_nHFQH19,Matsumoto_nHToric_PRL20,Guo_nHToric_EPL2020,Yoshida_nHFQHJ_PRR20,Zhang_nHToric_Natcomm20,Shackleton_nHFracton_PRR20,Zhang_nHTMI_arXiv20,Liu_nHTMI_RPB20,Xu_nHBM_arXiv20,Pan_PTHubb_oQS_arXiv20,Yang_EPKitaev_PRL21}.
For instance, it was reported that correlations induce topological ordered states such as fractional quantum Hall states~\cite{Yoshida_nHFQH19,Yoshida_nHFQHJ_PRR20} and spin liquid states~\cite{Matsumoto_nHToric_PRL20,Guo_nHToric_EPL2020,Zhang_nHToric_Natcomm20}. 
In addition, a previous work addressed classification of symmetry-protected topological phases with the non-trivial line-gap topology which have Hermitian counter parts~\cite{Xi_nHSPT_arXiv19}. This result implies that the reduction of topological classification of the line-gap topology should be observed as is the case of Hermitian systems. 
For Hermitian cases, the presence of correlations allow to adiabatically connect topological phases distinguished by a $\mathbb{Z}$-invariant in the non-interaction case~\cite{Z_to_Zn_Fidkowski_10,YaoRyu_Z_to_Z8_2013,Ryu_Z_to_Z8_2013,Qi_Z_to_Z8_2013}; for instance, a topological phase in eight copies of the Hermitian Kitaev chain can be adiabatically connected with maintaining the relevant symmetry~\cite{Z_to_Zn_Fidkowski_10}, which indicates the reduction of topological classification $\mathbb{Z}\to\mathbb{Z}_8$.

Unfortunately, however, correlation effects on the point-gap topology, which is a unique topological structure for non-Hermitian systems, have not been sufficiently addressed yet~\cite{Mu_MbdySkin_PRB20,Lee_MbdySkin_arXiv20}.
In particular, one may ask whether correlation effects induce the reduction of topological classification of the point-gap topology.
In addition, one may also ask whether interactions result in any unique topological phenomenon of the point-gap topology for correlated systems.

In this paper, we address the above questions by analyzing a zero-dimensional correlated system with chiral symmetry. 
Our analysis elucidates that correlations destroy an EP which separates two distinct topological phases in the non-interacting case; one of them is characterized by the zero-th Chern number $N_{0\mathrm{Ch}}=0$, and the other is characterized by $N_{0\mathrm{Ch}}=2$.
This result indicates that the presence of correlations allows to adiabatically connect the two distinct topological phases characterized by the zero-th Chern number without closing the point-gap, which is interpreted as the reduction of the $\mathbb{Z}$-classification of the point-gap topology, $\mathbb{Z}\to\mathbb{Z}_2$.
Furthermore, we also discover a Mott exceptional point (MEP), a unique EP for correlated systems, where only spin degrees of freedom are involved.

The rest of this paper are organized as follows. 
In Sec.~\ref{sec: h}, we briefly review a zero-dimensional one-body Hamiltonian with chiral symmetry. 
In Sec.~\ref{sec: model and symm}, we introduce a non-Hermitian many-body Hamiltonian~\cite{many-bodyH_ftnt} with correlations which preserves chiral symmetry. 
In Sec~\ref{sec: spec many-body H}, we demonstrate that topological phase transition points accompanied by EPs disappear in the presence of correlations, which indicates the reduction of $\mathbb{Z}$-classification of the point-gap topology. 
We show that correlations induce the MEP in Sec.~\ref{sec: MottEP} which is accompanied by a short summary.
The appendices are devoted to details of spectral flows of the non-Hermitian many-body Hamiltonian and a brief review of the reduction of topological classification for a zero-dimensional Hermitian system.

\section{
One-body Hamiltonian
}
\label{sec: h}

Consider a two-orbital system in zero dimension whose non-interacting Hamiltonian is written as
\begin{eqnarray}
h&=& \left(
\begin{array}{cc}
i\alpha & \beta \\
 \beta & -i\alpha
\end{array}
\right),
\end{eqnarray}
with real numbers $\alpha$ and $\beta$.
The off-diagonal term describes hybridization between orbital $A$ and $B$.

\begin{figure}[!h]
\begin{minipage}{0.475\hsize}
\begin{center}
\includegraphics[width=1\hsize,clip]{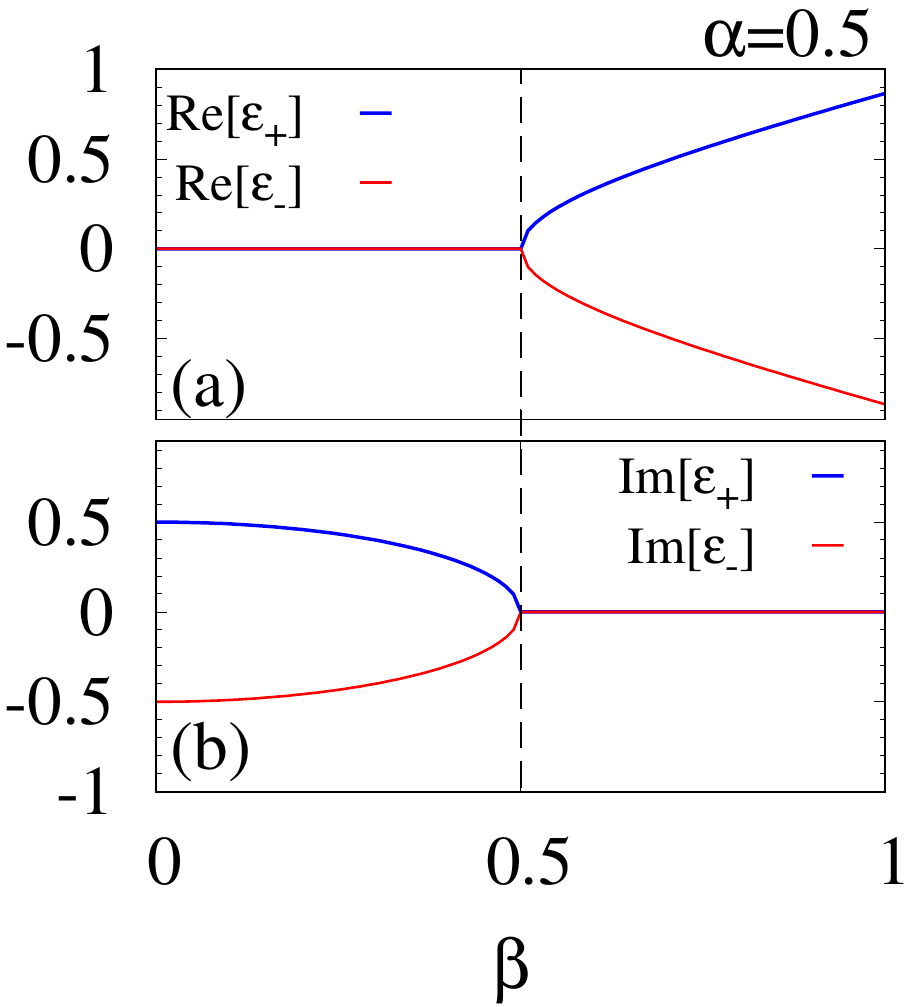}
\end{center}
\end{minipage}
\begin{minipage}{0.4\hsize}
\begin{center}
\includegraphics[width=1\hsize,clip]{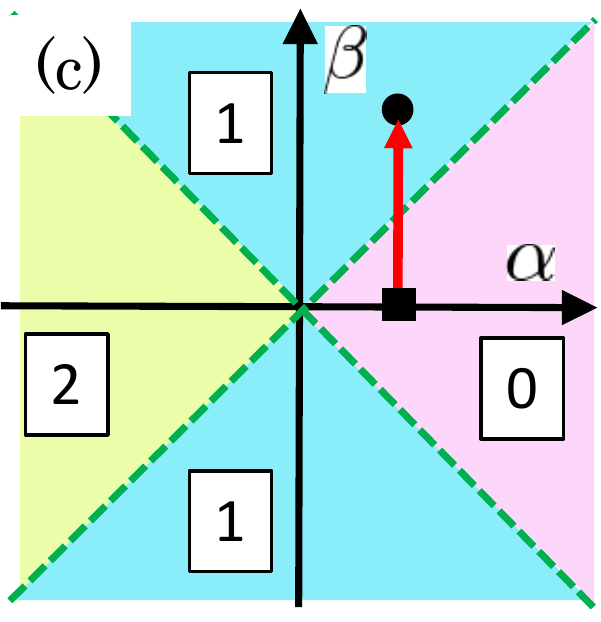}
\end{center}
\end{minipage}
\caption{
(a) [(b)]: The real and imaginary parts of $\epsilon_+$ and $\epsilon_-$ as functions of $\beta$ for $\alpha=0.5$.
(c): Phase diagram of the one-body Hamiltonian $h$. The Hamiltonian $h$ is characterized by the zero-th Chern number which is denoted by numbers in squares.
For $\alpha=\beta$ and $\alpha=-\beta$, the zero-th Chern number jumps. 
The black square (dot) in this panel denotes the point $(\alpha,\beta)=(0.5,0)$ [$(0.5,1)$].
With increasing $\beta$ along with the red arrow, the system shows an EP at $\alpha=\beta=0.5$ [see panels~(a)~and~(b)].
}
\label{fig: phase free}
\end{figure}

This model preserves chiral symmetry which satisfies
\begin{eqnarray}
\label{eq: chiral h}
\tau_3h^\dagger \tau_3&=&-h,
\end{eqnarray}
where $\tau$'s denote the Pauli matrices.
This relation imposes the following constraint on the spectrum of $h$:
the eigenvalues $\epsilon_+$ and $\epsilon_-$ are pure-imaginary or form a pair, $\epsilon_-=-\epsilon^*_+$. 
Namely, the symmetry requires the spectrum to be symmetric about the imaginary axis.
In Figs.~\ref{fig: phase free}(a)~and~\ref{fig: phase free}(b), eigenvalues are plotted where we can see that the eigenvalues satisfy the above constraint.

We note that for $\alpha\neq\beta$, the point-gap at $\epsilon_{\mathrm{ref}}=0$ opens; no eigenvalue is equal to $\epsilon_{\mathrm{ref}}=0$, where $\epsilon_{\mathrm{ref}}$ denotes the reference energy.
In this case, the point-gap topology is characterized by the zero-th Chern number which is the number of eigenstates with the negative eigenvalue of
\begin{eqnarray}
-i\tau_3h&=& \alpha\tau_0+\beta\tau_2.
\end{eqnarray}
Here, $\tau_0$ is the identity matrix.
We note that $N_{0\mathrm{Ch}}$ takes an arbitrary integer in the generic case, while it takes $0$, $1$, or $2$ for this model.
The phase diagram is plotted in Fig.~\ref{fig: phase free}(c).
For $(\alpha,\beta)=(0.5,0)$, the zero-th Chern number takes $N_{0\mathrm{Ch}}=0$. Increasing $\beta$, the zero-th Chern number jumps from $N_{0\mathrm{Ch}}=0$ to $1$ for $\beta=\alpha=0.5$ [see the red line illustrated in Fig.~\ref{fig: phase free}(c)]. 
Correspondingly, the point-gap at $\epsilon_{\mathrm{ref}}=0$ closes on this topological transition point due to the emergence of an EP.

In the above, we have seen the EP on the topological transition point which separates the two phases of $h$ characterized by $N_{0\mathrm{Ch}}$; one of them is characterized by $ N_{0\mathrm{Ch}}=0$ and the other is characterized by $ N_{0\mathrm{Ch}}=1$ [see Fig.~\ref{fig: phase free}(c)].
This fact indicates the $\mathbb{Z}$-classification of the point-gap topology with chiral symmetry.

\section{
Many-body Hamiltonian
}
\label{sec: model and symm}

\subsection{
Correlated model
}
\label{sec: model}

Now, let us consider the following model with correlations
\begin{subequations}
\label{eq: defs many-body H}
\begin{eqnarray}
\hat{H}&=& \hat{H}_{0}+\hat{H}_{\mu}+\hat{H}_{U},
\end{eqnarray}
\begin{eqnarray}
\hat{H}_0 &=& \hat{\Psi}^\dagger 
\left(
\begin{array}{cc}
h & 0 \\
0 & rh
\end{array}
\right)
\hat{\Psi},
\end{eqnarray}
\begin{eqnarray}
\hat{\Psi} &=& (\hat{c}_{A\uparrow},\hat{c}_{B\uparrow},\hat{c}_{A\downarrow},\hat{c}_{B\downarrow})^T,\\
\hat{H}_{\mu} &=& i\mu (\hat{N}_\uparrow-1)+ir\mu (\hat{N}_\downarrow-1),\\
\hat{H}_{U} &=& U \sum_{l=A,B} \left(\hat{n}_{l\uparrow}-\frac{1}{2}\right)\left(\hat{n}_{l\downarrow}-\frac{1}{2}\right).
\end{eqnarray}
\end{subequations}
Here, $\hat{c}^\dagger_{ls}$ ($\hat{c}_{ls}$) creates (annihilates) a fermion in orbital $l=A,B$ and spin state $s=\uparrow,\downarrow$.
The number operator of up-spin states (down-spin states) is defined as $\hat{N}_\uparrow=\sum_l \hat{n}_{l\uparrow}$ ($\hat{N}_\downarrow=\sum_l \hat{n}_{l\downarrow}$) with $\hat{n}_{ls}=c^\dagger_{ls}c_{ls}$.
Interaction strength is described by a real number $U\geq 0$, and $\mu$ is a real number.
The factor $r$ ($0<r\leq 1$) is introduced for a technical reason~\cite{r_ftnt}.
We consider that the above Hamiltonian is relevant to open quantum systems with one-body loss under continuous observations.

For $U=0$, the Hamiltonian is decomposed to 
\begin{eqnarray}
\label{eq: H=Hup Hdow}
\hat{H}_{0}+\hat{H}_{\mu}&=&\hat{H}_{0\uparrow}+\hat{H}_{0\downarrow}.
\end{eqnarray}
Here, $\hat{H}_{0\uparrow}$ ($\hat{H}_{0\downarrow}$) acts only on fermions in up-spin states (down-spin states).

\subsection{
Symmetry of the many-body Hamiltonian
}
\label{sec: symm}

The Hamiltonian is chiral symmetric 
\begin{subequations}
\label{eq: many-body chiral}
\begin{eqnarray}
\hat{\Xi} \hat{H} \hat{\Xi}^{-1}&=& \hat{H},
\end{eqnarray}
\begin{eqnarray}
\hat{\Xi} &=& \prod_s(\hat{c}^\dagger_{As}+\hat{c}_{As})(\hat{c}^\dagger_{Bs}-\hat{c}_{Bs})\mathcal{K},
\end{eqnarray}
\end{subequations}
with operator $\mathcal{K}$ taking complex conjugate.
Here, $\hat{\Xi}$ is written as a product of a time-reversal operator and a particle-hole operator.

For $\mu=U=0$, the chiral symmetry of the many-body Hamiltonian is reduced to Eq.~(\ref{eq: chiral h}), which can be confirmed as follows~\cite{Hatsugai_chiralop_JPSJ06,Gurarie_chiral_PRB11}.
Noting the relation
\begin{eqnarray}
\hat{\Xi} \hat{\Psi} \hat{\Xi}^{-1} &=& 
\left(
\begin{array}{cc}
\tau_3 &  \\
 & \tau_3
\end{array}
\right)
\hat{\Psi}^*,
\end{eqnarray}
with 
\begin{eqnarray}
\hat{\Psi}^* &=& 
(\hat{c}^\dagger_{A\uparrow}, \hat{c}^\dagger_{B\uparrow}, \hat{c}^\dagger_{A\downarrow}, \hat{c}^\dagger_{B\downarrow})^T,
\end{eqnarray}
we obtain
\begin{eqnarray}
\hat{\Xi} \hat{H}_{0} \hat{\Xi}^{-1} &=& 
\hat{\Psi}^T
\left(
\begin{array}{cc}
\tau_3 h^*\tau_3 & 0 \\
0 & \tau_3 rh^* \tau_3
\end{array}
\right)
\hat{\Psi}^*,
\nonumber \\
&=&
\hat{\Psi}^\dagger
\left(
\begin{array}{cc}
-\tau_3 h^\dagger \tau_3 & 0 \\
0 & -\tau_3 rh^\dagger \tau_3
\end{array}
\right)
\hat{\Psi}.
\end{eqnarray}
Here, from the first to the second line, we have used the fact that $h$ is a traceless matrix. The minus sign in the second line arises from the fermionic statistics.
Thus, for $U=0$, Eq.~(\ref{eq: many-body chiral}) is reduced to Eq.~(\ref{eq: chiral h}), indicating that $\hat{H}_0$ satisfies $\hat{\Xi} \hat{H}_0 \hat{\Xi}^{-1}=\hat{H}_0$ when $h$ is chiral symmetric.

We note that $\hat{H}_\mu$ and $\hat{H}_U$ are also chiral symmetric which can be seen by noting the relation
\begin{eqnarray}
\hat{\Xi} \hat{n}_{ls} \hat{\Xi}^{-1}&=& 1-\hat{n}_{ls}.
\end{eqnarray}
Thus, the many-body Hamiltonian is chiral symmetric for arbitrary values of $\mu$ and $U$ [see Eq.~(\ref{eq: many-body chiral})].

The eigenvalues $E_n$ ($n=1,2,\ldots$) of the many-body Hamiltonian satisying Eq.~(\ref{eq: many-body chiral}) are real or form pairs satisfying
\begin{eqnarray}
\label{eq: En=E^*np}
E_{n'}&=& E^*_{n}.
\end{eqnarray}
%
Here, we note a crucial difference between symmetry constraints Eqs.~(\ref{eq: chiral h})~and~(\ref{eq: many-body chiral}).
As seen in Sec.~\ref{sec: h}, the constraint on the one-body Hamiltonian [Eq.~(\ref{eq: chiral h})] requires the spectrum of $h$ to be symmetric about the imaginary axis.
In contrast, the constraint on the many-body Hamiltonian requires the spectrum of $\hat{H}$ to be symmetric about the real axis [see Eq.~(\ref{eq: En=E^*np})].
The difference arises from the fact that the chiral symmetry of the many-body Hamiltonian [Eq.~(\ref{eq: many-body chiral})] is mathematically the same as the time-reversal symmetry; we note that the operator of the particle-hole transformation is unitary in the second quantized form.

We also note that the Hamiltonian $\hat{H}$ commutes with $\hat{N}_{\mathrm{tot}}=\sum_{ls}\hat{n}_{ls}$ and $\hat{S}^z_{\mathrm{tot}}= (\hat{N}_\uparrow-\hat{N}_\downarrow)/2$.

\section{
Reduction of topological classification of point-gap topology
}
\label{sec: spec many-body H}

We show that the EP discussed in Sec.~\ref{sec: h} vanishes due to the correlation effects. 
This fact indicates that correlations allow to continuously connect two distinct topological phases of free fermions with maintaining the point-gap at $E_{\mathrm{ref}}=0$; the one of them is characterized by $N_{0\mathrm{Ch}}=0$, and the other is characterized by $N_{0\mathrm{Ch}}=2$.

\subsection{
Case for $U=0$
}
\label{sec: spec of many H U=0}

Let us start with the spectrum for $U=0$ which can be obtained from eigenvalues of $h$.
We recall that the Hamiltonian can be block-diagonalized with operators $\hat{N}_{\mathrm{tot}}$ and $\hat{S}^z_{\mathrm{tot}}$.

\begin{figure}[!h]
\begin{minipage}{1\hsize}
\begin{center}
\includegraphics[width=1\hsize,clip]{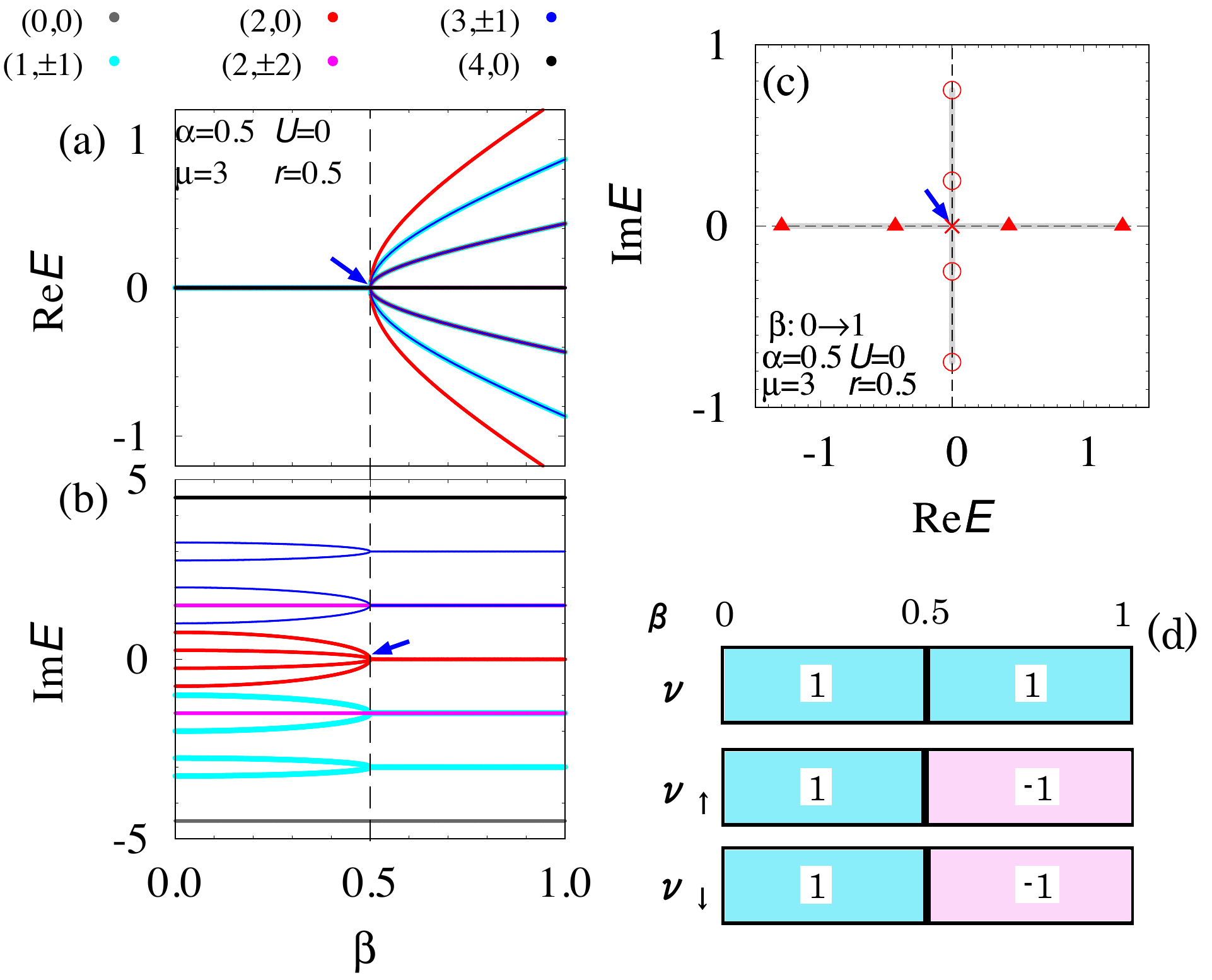}
\end{center}
\end{minipage}
\caption{
(a) [(b)]: The real and imaginary parts of eigenvalues as functions of $\beta$. Colors denote eigenvalues for subsectors with $(\hat{N}_{\mathrm{tot}},2\hat{S}^z_{\mathrm{tot}})$.
(c): Spectral flow for subsector with $(\hat{N}_{\mathrm{tot}},2\hat{S}^z_{\mathrm{tot}})=(2,0)$. Circles, crosses, and triangles denote eigenvalues for $\beta=0$, $0.5$, and $1$, respectively.
With increasing $\beta$ from $0$ to $1$, eigenvalues flow along gray lines.
(d): $\mathbb{Z}_2$-invariants as functions of $\beta$. At $\alpha=\beta=0.5$, the point-gap at $E_{\mathrm{ref}}=0$ closes.
These data are obtained for $\alpha=0.5$, $U=0$, $\mu=3$, and $r=0.5$
}
\label{fig: spectrum U=0}
\end{figure}

Figures~\ref{fig: spectrum U=0}(a)~and~\ref{fig: spectrum U=0}(b) plot the spectrum as a function of $\beta$ for $\alpha=0.5$. 
Here, we set $\mu$ to $3$ ($\mu=3$) in order to focus on the topology of subsector with $(\hat{N}_{\mathrm{tot}},2\hat{S}^z_{\mathrm{tot}})=(2,0)$.
In these figures, we can see that the system closes the point-gap at $E_{\mathrm{ref}}=0$ by showing an EP at zero-energy [see blue arrows]. 
Here, $E_{\mathrm{ref}}=0$ denotes the reference energy.
This EP is also observed in Fig.~\ref{fig: spectrum U=0}(c) where a spectral flow~\cite{U0spec_awyhf_comm_ftnt} for subsector with $(\hat{N}_{\mathrm{tot}},2\hat{S}^z_{\mathrm{tot}})=(2,0)$ is plotted.
At $\beta=0$, energy eigenvalues are pure-imaginary as denoted by open circles. 
Increasing $\beta$, we can see that the EP emerges at $\beta=0.5$ as denoted by the cross. 
At $\beta=1$, energy eigenvalues are real as denoted by triangles.

This EP of the many-body Hamiltonian $\hat{H}$ is induced by the topological transition of the one-body Hamiltonian $h$.
Namely, as discussed in Sec.~\ref{sec: h}, the zero-th Chern number $N_{0\mathrm{Ch}}$ jumps from $0$ to $1$ for $\alpha=\beta=0.5$. 
Correspondingly, the point-gap of $h$ at $\epsilon_{\mathrm{ref}}=0$ closes, which inducing the EP [see Figs.~\ref{fig: phase free}(a)~and~\ref{fig: phase free}(b)].
We also note that $\hat{H}_{\mu}$ is zero for subsector with $(\hat{N}_{\mathrm{tot}},2\hat{S}^z_{\mathrm{tot}})=(2,0)$.

For the other sectors, the system shows EPs at $E=\pm3i$ and $\pm1.5i$ [see Figs.~\ref{fig: spectrum U=0}(a)~and~\ref{fig: spectrum U=0}(b)], which can also be understood in terms of $h$ because $\hat{H}_{\mu}$ is reduced to a constant value for each sector~\cite{N1N3_free_ftnt} specified by $\hat{N}_{\mathrm{tot}}$ and $\hat{S}^z_{\mathrm{tot}}$.
As seen in Appendix~\ref{sec: all spec app}, the spectrum is symmetric about the real axis for $0 \leq  \beta \leq 1$ due to the chiral symmetry [Eq.~(\ref{eq: many-body chiral})].

In the above, we have seen the emergence of the EP at $E=0$ which separates two distinct topological phases; one of them is characterized by $N_{0\mathrm{Ch}}=0$, and the other is characterized by $N_{0\mathrm{Ch}}=2$. Here, we have taken into account spin degrees of freedom.

\subsection{
Case for $U>0$
}
\label{sec: spec of many H U=1}

We show that the EP at $E=0$ is fragile against the interaction.
This fragility of the EP implies the reduction of topological classification of the point-gap topology $\mathbb{Z}\to\mathbb{Z}_2$; correlations allow to continuously connect the topological phase characterized by $N_{0\mathrm{Ch}}=0$ and the one characterized by $N_{0\mathrm{Ch}}=2$ without closing the point-gap at $E_{\mathrm{ref}}=0$.

As a first step to understand the fragility of the EP at $E=0$, we recall that the symmetry constraint of the many-body Hamiltonian is mathematically equivalent to that of the time-reversal symmetry [see Eq.~(\ref{eq: many-body chiral})], which differs from the constraint on the one-particle Hamiltonian $h$ [see Eq.~(\ref{eq: chiral h})].
This fact indicate that the point-gap topology of the many-body Hamiltonian is characterized by the following $\mathbb{Z}_2$-invariant~\cite{Gong_class_PRX18}
\begin{eqnarray}
\label{eq: many-body Z2}
\nu &=& \sgn [\mathrm{det} (\hat{H}-E_{\mathrm{ref}})],
\end{eqnarray}
with the reference energy $E_{\mathrm{ref}}\in \mathbb{R}$.
In the non-interacting case, we can see that the two pairs of energy eigenvalues touch at $E=0$ for $\alpha=\beta$ with increasing $\beta$ [see Figs.~\ref{fig: spectrum U=0}(a)-\ref{fig: spectrum U=0}(c)].
Correspondingly, for $E_{\mathrm{ref}}=0$, topological invariants $\nu_{\uparrow}$ and $\nu_{\downarrow}$ jump from $1$ to $-1$ [see Fig.~\ref{fig: spectrum U=0}(d)], where $\nu_{\uparrow}$ ($\nu_{\downarrow}$) is defined by replacing $\hat{H}$ to $\hat{H}_{0\uparrow}$ ($\hat{H}_{0\downarrow}$) in Eq.~(\ref{eq: many-body Z2})~\cite{HupHdow_ftnt}.
Computing the $\mathbb{Z}_2$-invariant $\nu$ for $U=0$, we can see that $\nu$ does not change its value for $0 \leq \beta \leq 1$ [see Fig.~\ref{fig: spectrum U=0}(d)], although the EP at $E=0$ is observed for $\alpha=\beta=0.5$ [see Fig.~\ref{fig: spectrum U=0}].
Noting that for $U>0$, $\nu$ is the only topological invariant~\cite{decoupledspin_ftnt} among $\nu$, $\nu_\uparrow$, and $\nu_\downarrow$, we can conclude that the EP at $E=0$ is fragile against correlations.

The fragility of the EP at $E=0$ is also observed by computing spectrum of the many-body Hamiltonian. 
\begin{figure}[!h]
\begin{minipage}{1\hsize}
\begin{center}
\includegraphics[width=1\hsize,clip]{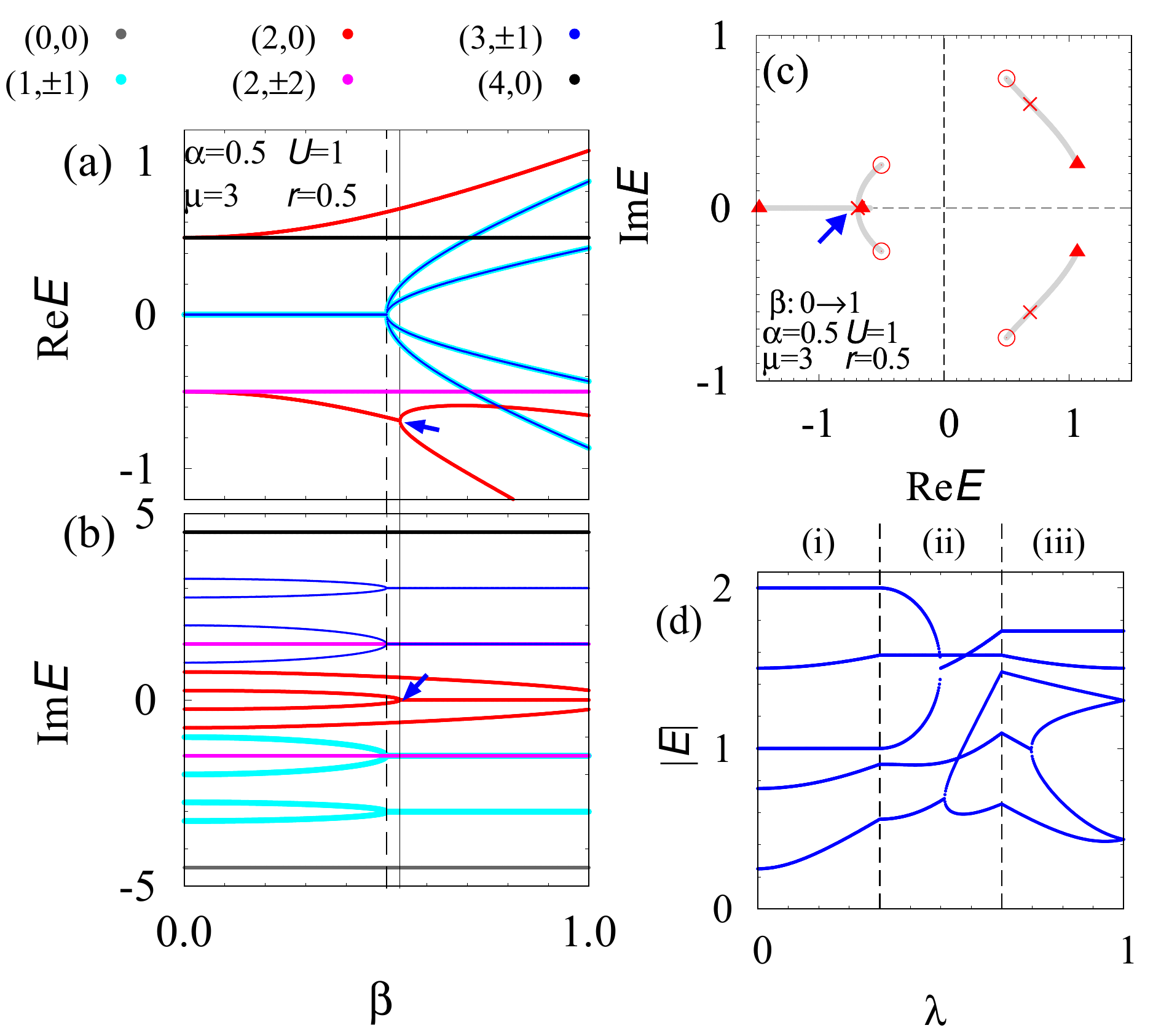}
\end{center}
\end{minipage}
\caption{
(a) [(b)]: The real and imaginary parts of eigenvalues as functions of $\beta$. Colors denote eigenvalues for subsectors with $(\hat{N}_{\mathrm{tot}},2\hat{S}^z_{\mathrm{tot}})$.
(c): Spectrum for subsector with $(\hat{N}_{\mathrm{tot}},2\hat{S}^z_{\mathrm{tot}})=(2,0)$. Circles, crosses, and triangles denote eigenvalues for $\beta=0$, $0.533$, and $1$, respectively.
With increasing $\beta$ from $0$ to $1$, eigenvalues flow along gray lines.
The spectral flow including other sectors is shown in Appendix~\ref{sec: all spec app}.
In panels~(a),~(b),~and~(c), blue arrows denote an EP away from $E=0$ which cannot be described by the one-body Hamiltonian $h$.
(d): The absolute value of eigenvalues as functions of $\lambda$. 
The reference energy is $|E_{\mathrm{ref}}|=0$. 
Here, $\lambda$ parameterizes a deformation as follows: (i) for $0\leq \lambda<1/3$, it parameterizes as $(U,\beta)=(3\lambda,0)$;
(ii) for $1/3\leq \lambda<2/3$, it parameterizes as $(U,\beta)=(1,3\lambda-1)$;
(iii) for $2/3\leq \lambda<1$, it parameterizes as $(U,\beta)=(3-3\lambda,1)$.
In panel (d), dashed vertical lines denote $\lambda=1/3$ and $\lambda=2/3$, respectively.
These data are obtained for $\alpha=0.5$, $\mu=3$, and $r=0.5$.
}
\label{fig: spectrum U1}
\end{figure}
Figures~\ref{fig: spectrum U1}(a)-\ref{fig: spectrum U1}(c) elucidate the absence of the EPs at $E=0$ which are observed in the non-interacting case~\cite{briefintroMottEP_ftnt}; although we observe an EP for $U=1$, it emerges away from $E=0$.
These results are consistent with the $\mathbb{Z}_2$-invariant $\nu$; 
the point-gap at $E_{\mathrm{ref}}=0$ remains open for $\alpha=0.5$ and $ 0 \leq  \beta \leq 1$, corresponding to the fact that $\nu$ does not change its value.

In addition, Fig.~\ref{fig: spectrum U1}(d) indicate that the point-gap at $E_{\mathrm{ref}}=0$ also remains open under the following deformation: increasing $U$ from $0$ to $1$ for $(\alpha,\beta)=(0.5,0)$; decreasing $U$ from $1$ to $0$ for $(\alpha,\beta)=(0.5,1)$.

The above results [see Figs.~\ref{fig: spectrum U1}(c)~and~\ref{fig: spectrum U1}(d)] indicate that correlation effects allow to connect two topological phases for $(\alpha,\beta)=(0.5,0)$ and $(0.5,1)$ without closing the point-gap at $E_{\mathrm{ref}}=0$. The former (latter) is characterized by $N_{0\mathrm{Ch}}=0$ ($N_{0\mathrm{Ch}}=2$) in the non-interacting case.
This behavior is reminiscent of the reduction of topological classification; for instance, in a zero-dimensional topological system, the topology of the one-body Hamiltonian is characterized by a $\mathbb{Z}$-invariant while the topology of two copies of them are fragile against the interactions (for more details of the Hermitian case, see Appendix~\ref{sec: reduction Hermi app}).
We would like to stress that the interaction is essential for this phenomenon as is the case of a Hermitian system~\cite{Z_to_Zn_Fidkowski_10}. Namely, without any interaction term, one cannot avoid the topological transition.  

\section{
Mott exceptional point
}
\label{sec: MottEP}

In the above, we have seen that the point-gap at $E_{\mathrm{ref}}=0$ remains open while the EP emerges away from $E=0$.
In this section, we show that this EP corresponds to the MEP, a unique EP for correlated systems, where only spin degrees of freedom are involved.

Firstly, we recall that for the non-interacting case, the EPs are fixed to the imaginary axis [see Figs.~\ref{fig: spectrum U=0}(a),~\ref{fig: spectrum U=0}(b),~and~\ref{fig: Eall U0 app}] because the eigenvalues are governed by the one-body Hamiltonian $h$.
In contrast to these EPs for $U=0$, the MEP emerges away from the imaginary axis [see Fig.~\ref{fig: Eall U1 app}(c)], which indicates that interactions are essential.

For better understanding of this MEP, we apply the second order perturbation theory by supposing that the interaction $U$ is sufficiently large $U\gg \alpha,\beta$.
In such a case, an EP emerges around $\mathrm{Re}E=-U/2$. At $\alpha=\beta=0$, this band touching is described by the following two states:
\begin{eqnarray}
| \Psi_{1} \rangle &=& \hat{c}^\dagger_{A\uparrow} \hat{c}^\dagger_{B\downarrow} |0\rangle, \\
| \Psi_{2} \rangle &=& \hat{c}^\dagger_{A\downarrow} \hat{c}^\dagger_{B\uparrow} |0\rangle.
\end{eqnarray}
For this subspace, we can obtain the effective Hamiltonian
\begin{eqnarray}
\hat{H}_{\mathrm{eff}} &=& -\frac{U}{2} +i\alpha (1-r) \sum_l \sgn(l) \hat{S}^z_{l} +\frac{(1+r^2)}{2}J \hat{S}^z_A \hat{S}^z_B \nonumber \\ 
                  && +rJ(\hat{S}^x_A \hat{S}^x_B+\hat{S}^y_A \hat{S}^y_B) -\frac{J(1+r^2)}{8},
\end{eqnarray}
with $J=4\beta^2/U$.
Here, $\hat{S}^\mu_{l}$ ($\mu=x,y,z$) are spin operators; $\hat{S}^z_{l}=(\hat{n}_{l\uparrow}-\hat{n}_{l\downarrow})/2$, $\hat{S}^x_{l}=(\hat{S}^+_l+\hat{S}^-_l)/2$, and $\hat{S}^y_{l}=(\hat{S}^+_l-\hat{S}^-_l)/(2i)$ 
with $\hat{S}^{+}_{l}=\hat{c}^\dagger_{l\uparrow}\hat{c}_{l\downarrow}$ and $\hat{S}^{-}_{l}=\hat{c}^\dagger_{l\downarrow}\hat{c}_{l\uparrow}$.
In a matrix form, the effective Hamiltonian and the operator $\hat{\Xi}$ are rewritten as
\begin{subequations}
\label{eq: Heff}
\begin{eqnarray}
\hat{H}_{\mathrm{eff}}&=& E_0\rho_0 + \frac{rJ}{2}\rho_1 +i\alpha(1-r)\rho_3,\\
\hat{\Xi} &=& \rho_1\mathcal{K},
\end{eqnarray}
\end{subequations}
with $E_0=-U/2-J(1+r^2)/4$.
Here, $\rho_0$ and $\rho$'s are the identity matrix and the Pauli matrices, respectively.
Diagonalizing the Hamiltonian, we have 
\begin{eqnarray}
E&=& E_0 \pm \sqrt{(\frac{rJ}{2})^2-\alpha^2(1-r)^2}.
\end{eqnarray}
Thus, for $J=J_{c}:=2\alpha(1-r)/r$, we have an EP. 
Here, we would like to stress that the above EP is described by the effective spin model~(\ref{eq: Heff}), which
indicates that correlations are essential for the MEP.

\begin{figure}[!h]
\begin{minipage}{0.7\hsize}
\begin{center}
\includegraphics[width=1\hsize,clip]{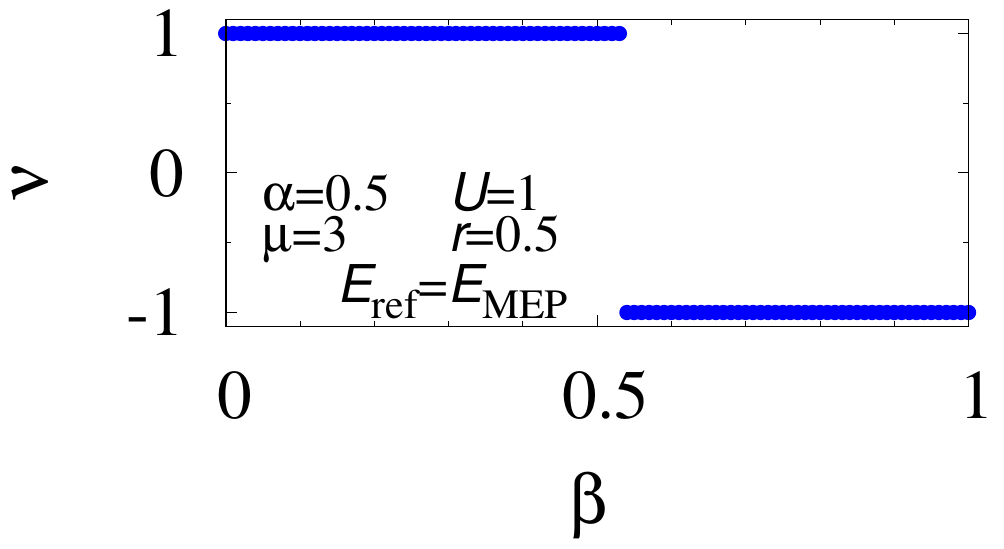}
\end{center}
\end{minipage}
\caption{
The $\mathbb{Z}_2$-invariant as a function of $\beta$. Here, the reference energy is set to $E_{\mathrm{ref}}=E_{\mathrm{MEP}}=-0.6886$ where the MEP emerges.
The data are obtained for $\alpha=0.5$, $U=1$, $\mu=3$, and $r=0.5$.
}
\label{fig: nu Mott}
\end{figure}
Because $H_{\mathrm{eff}}$ preserves the chiral symmetry, we can characterize the MEP by the $\mathbb{Z}_2$-invariant
\begin{eqnarray}
\label{eq: nu eff}
\nu_{\mathrm{eff}} &=& \mathrm{sgn}[\mathrm{det} ( H_{\mathrm{eff}}-E_0\rho_0)].
\end{eqnarray}
Here, we have fixed the reference energy to $E_{\mathrm{ref}}=E_0$.
By substituting Eq.~(\ref{eq: Heff}a) to Eq.~(\ref{eq: nu eff}), we obtain $\nu_{\mathrm{eff}}$ taking $1$ ($-1$) for $J<J_c$ ($J>J_c$).
This result is consistent with the $\mathbb{Z}_2$-invariant computed from the original Hamiltonian $\hat{H}$.
Figure~\ref{fig: nu Mott} shows that the $\mathbb{Z}_2$-invariant jumps at $\beta_c\sim0.53$, which indicates that the invariant jumps at $J \sim 1.12$. 

We note that the difference of the MEP from the ordinary EP for $U=0$ can also be seen by turning off $\mu$ (see Appendix~\ref{sec: all spec mu 0 app}).

\section{
Summary and discussion
}

In this paper, we have analyzed correlation effects on the point-gap topology in the zero-dimensional system with chiral symmetry.
Our analysis elucidates that correlations destroy the EP which separates two distinct topological phases characterized by the zero-th Chern number; one of them is characterized by $N_{0\mathrm{Ch}}=0$ and the other is characterized by $N_{0\mathrm{Ch}}=2$.
This result originates from the fact that the many-body chiral symmetry results in the $\mathbb{Z}_2$-invariant, in contrast to the $\mathbb{Z}$-invariant in the non-interacting case.
The above results suggest that correlations change $\mathbb{Z}$-classification of free fermions to $\mathbb{Z}_2$-classification, which is reminiscent of the reduction of topological classification in Hermitian systems.
Furthermore, we have discovered the MEP for which correlations are essential. The MEP is described by the effective spin Hamiltonian (i.e., charge degrees of freedom are not involved).

We finish this paper with a remark on the case of higher dimensions. 
Because Eqs.~(\ref{eq: chiral h})~and~(\ref{eq: many-body chiral}) should hold also for higher dimensions, the topological properties of the many-body Hamiltonian may differ from those of the one-body Hamiltonian for even spatial dimensions.

\section*{
Acknowledgments
}
T.Y. thanks Yoshihito Kuno for fruitful discussion.This work is supported by JPSP Grant-in-Aid for Scientific Research on Innovative Areas ``Discrete Geometric Analysis for Materials Design”: Grants No. JP20H04627 (T.Y.).  This work is also supported by the JSPS KAKENHI, Grants No~JP17H06138 and No.~JP21K13850.


\begin{thebibliography}{89}
\bibitem [{man()}]{many-bodyH_ftnt}%
  \BibitemOpen
  \href@noop {} {}\bibinfo {note} {{ In zero dimension, systems are described
  by few particle problems. However, we use the word ``many-body Hamiltonian"
  in order to distinguish it from one-body Hamiltonian }}\BibitemShut {NoStop}%
\bibitem [{r_f()}]{r_ftnt}%
  \BibitemOpen
  \href@noop {} {}\bibinfo {note} {{ For $r=1$, we have eigenstates whose
  eigenvalues are zero. In order to have a gapped system for
  $\epsilon_{\mathrm{ref}}=0$, we set $r=0.5$. We note that for
  $\epsilon_{\mathrm{ref}}=0$, the point-gap topology of $rh$ is the same as
  that of $h$ }}\BibitemShut {NoStop}%
\bibitem [{U0s()}]{U0spec_awyhf_comm_ftnt}%
  \BibitemOpen
  \href@noop {} {}\bibinfo {note} {{ In Fig.~\ref{fig: spectrum U=0}(c),
  eigenvalues for subsector with $(\hat{N}_{\mathrm{tot}},
  2\hat{S}^z_{\mathrm{tot}})=(2,0)$ are plotted. The spectrum including the
  other sectors are plotted in Fig.~\ref{fig: Eall U0 app} }}\BibitemShut
  {NoStop}%
\bibitem [{Hup()}]{HupHdow_ftnt}%
  \BibitemOpen
  \href@noop {} {}\bibinfo {note} {{ We recall Eq.~(\ref{eq: H=Hup Hdow})
  indicating that the Hamiltonian $\hat{H}$ is decomposed to
  $\hat{H}_{0\uparrow}$ and $\hat{H}_{0\downarrow}$ for $U=0$ }}\BibitemShut
  {NoStop}%
\bibitem [{dec()}]{decoupledspin_ftnt}%
  \BibitemOpen
  \href@noop {} {}\bibinfo {note} {{ This is because up-spin and down-spin
  states are decoupled only for the non-interacting case. Only in the
  non-interacting case, the eigenvalues of $\hat{H}$ is obtained by summing the
  eigenvalues of $\hat{H}_{0\uparrow}$ and $\hat{H}_{0\downarrow}$
  }}\BibitemShut {NoStop}%
\bibitem [{bri()}]{briefintroMottEP_ftnt}%
  \BibitemOpen
  \href@noop {} {}\bibinfo {note} {{ These figures indicate the emergence of an
  EP away from $E=0$. For this EP, correlation effects is essential (see for
  more details, see Sec.~\ref{sec: MottEP}) }}\BibitemShut {NoStop}%
\end{thebibliography}
%


\appendix

\section{
Details of spectral flows
}
\label{sec: all spec app}
\begin{figure}[!h]
\begin{minipage}{1\hsize}
\begin{center}
\includegraphics[width=1\hsize,clip]{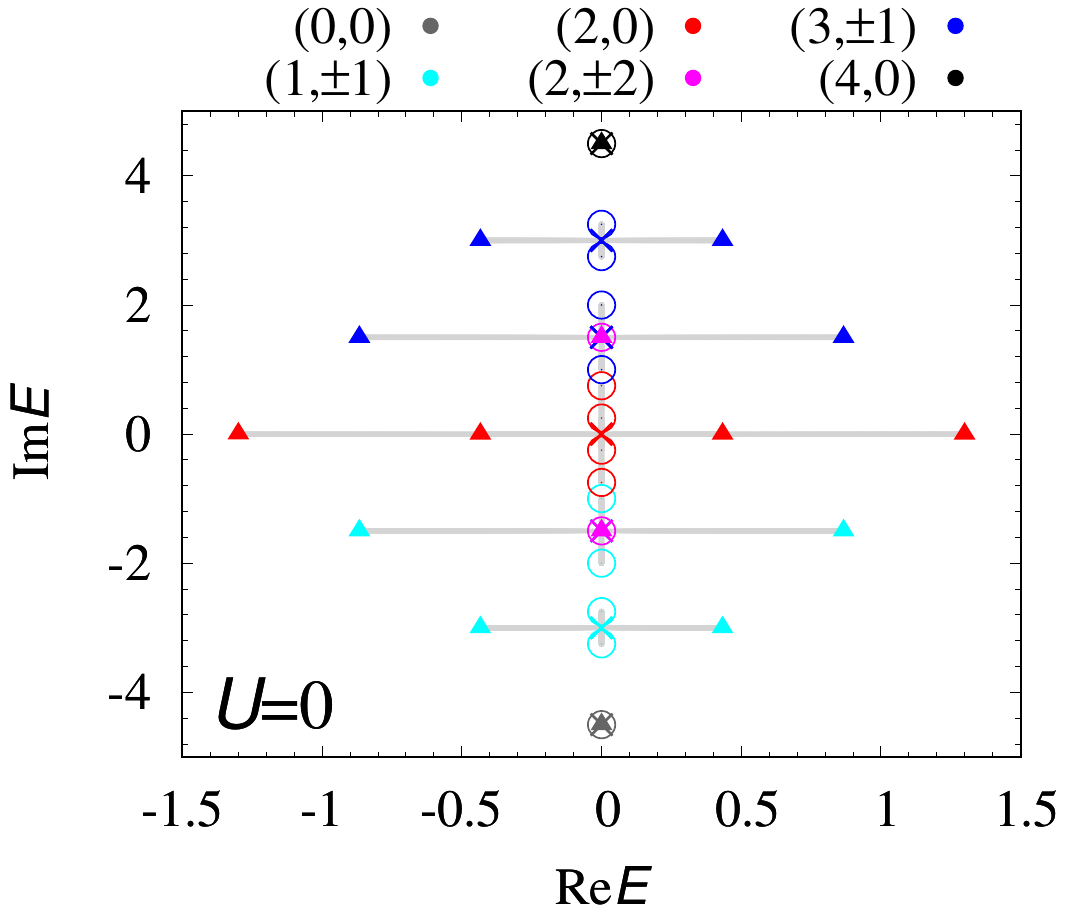}
\end{center}
\end{minipage}
\caption{
Spectral flow for each subsector with $(\hat{N}_{\mathrm{tot}},2\hat{S}^z_{\mathrm{tot}})$. Circles, crosses, and triangles denote eigenvalues for $\beta=0$, $0.5$, and $1$, respectively.
With increasing $\beta$ from $0$ to $1$, eigenvalues flow along gray lines.
These data are obtained for $U=0$, $\alpha=0.5$, $\mu=3$, and $r=0.5$
}
\label{fig: Eall U0 app}
\end{figure}

Here, we plot the spectral flow of $\hat{H}$ in the complex plane.
Figure~\ref{fig: Eall U0 app} indicates the spectral flow for $U=0$. 
In this case, with increasing $\beta$ from $0$ to $1$, we can see that a topological phase transition occurs at $\beta=\alpha$ [see also Fig.~\ref{fig: phase free}].
Correspondingly, an EP emerges, and the point-gap at $E_{\mathrm{ref}}=0$ closes for $\alpha=\beta$.

\begin{figure}[!h]
\begin{minipage}{1\hsize}
\begin{center}
\includegraphics[width=1\hsize,clip]{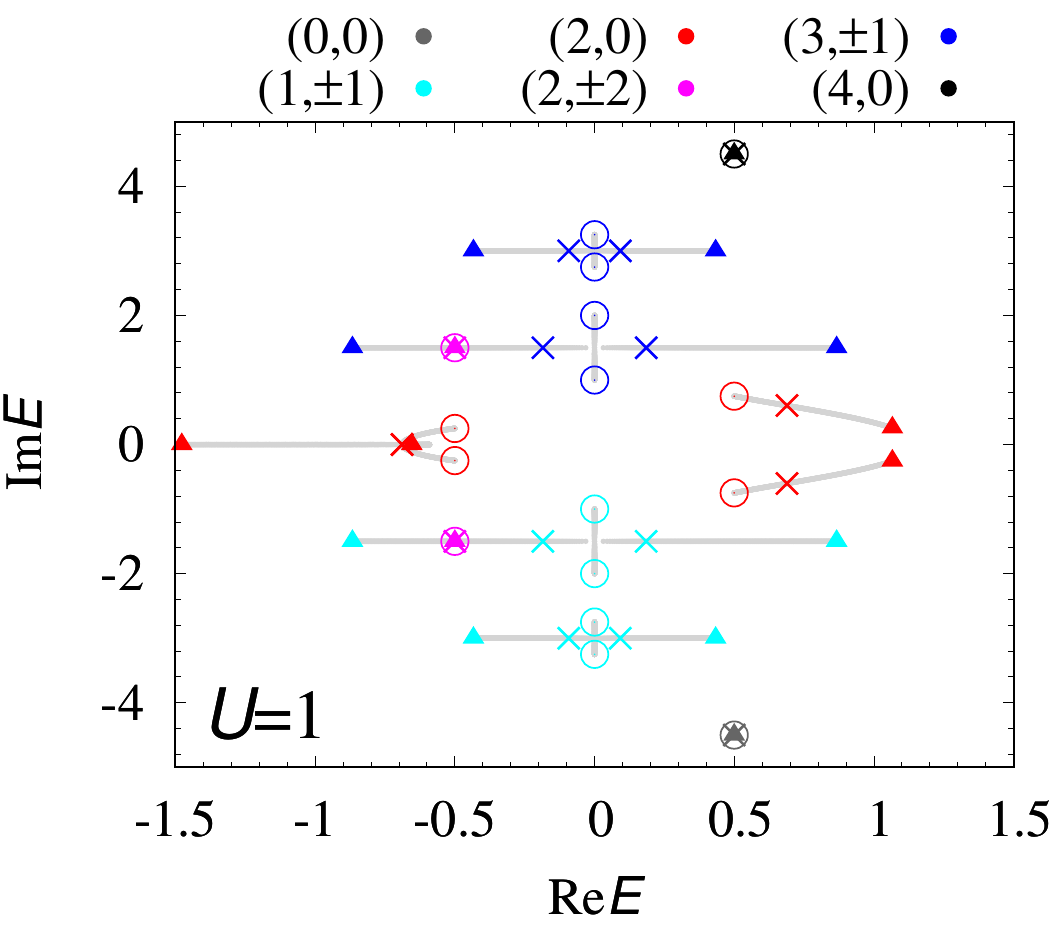}
\end{center}
\end{minipage}
\caption{
Spectral flow for each subsector with $(\hat{N}_{\mathrm{tot}},2\hat{S}^z_{\mathrm{tot}})$. Circles, crosses, and triangles denote eigenvalues for $\beta=0$, $0.533$, and $1$, respectively.
With increasing $\beta$ from $0$ to $1$, eigenvalues flow along gray lines.
These data are obtained for $U=1$, $\alpha=0.5$, $\mu=3$, and $r=0.5$.
We note that the eigenvalues denoted by red symbols are identical to the ones shown in Fig.~\ref{fig: spectrum U1}(c).
}
\label{fig: Eall U1 app}
\end{figure}

In contrast to the gap-closing observed for $U=0$, the point-gap at $E_{\mathrm{ref}}=0$ remains open for $U=1$.
Figure~\ref{fig: Eall U1 app} indicates the spectral flow for $U=1$. 
This figure shows that the point-gap at $E_{\mathrm{ref}}=0$ remains open for $0\leq \beta \leq 1$.

\section{
Reduction of topological classification for a Hermitian case
}
\label{sec: reduction Hermi app}

Let us briefly review reduction of topological classification for a Hermitian system in zero dimension.

Consider a zero-dimensional Hamiltonian $\hat{H}_{\mathrm{H}0}$ defined as
\begin{eqnarray}
\hat{H}_{\mathrm{H}0}&=& 
\hat{\Psi}^\dagger
\left(
\begin{array}{cc}
h_{\mathrm{H}} &  0 \\
0 & rh_{\mathrm{H}}
\end{array}
\right)
\hat{\Psi},
\end{eqnarray}
with $r$ ($0 \leq r \leq 1$) and $\hat{\Psi}$ defined in Eq.~(\ref{eq: defs many-body H}c).
When the Hamiltonian is invariant under applying $\hat{U}_{\mathrm{H}}=e^{i\pi\hat{T}^z_{\mathrm{tot}}}$ with $\hat{T}^z_{\mathrm{tot}}=\sum_{s}(\hat{n}_{as}-\hat{n}_{bs})/2$, the symmetry constraint $[\hat{H}_{\mathrm{H}0},\hat{U}_{\mathrm{H}}]=0$ is rewritten as
\begin{eqnarray}
[h_{\mathrm{H}},i\tau_3]&=& 0,
\end{eqnarray}
which can be seen by noting the following relation
\begin{eqnarray}
\label{eq: U psi Udag app}
\hat{U}_{\mathrm{H}} \hat{\Psi} \hat{U}^\dagger_{\mathrm{H}} &=& 
\left(
\begin{array}{cc}
i\tau_3 &  \\
 & i\tau_3
\end{array}
\right)
\hat{\Psi}.
\end{eqnarray}

Namely, the one-body Hamiltonian $h_{\mathrm{H}}$ is written as $h_{\mathrm{H}}=\alpha_0 \tau_0+\alpha\tau_3$ with real numbers $\alpha_0$ and $\alpha$.
In the non-interacting case, the topological phase is characterized by the $\mathbb{Z}$-invariant, $N_{\mathrm{ps0Ch}}=(N_{A0\mathrm{Ch}}-N_{B0\mathrm{Ch}})/2$, 
where $N_{A0\mathrm{Ch}}$ ($N_{B0\mathrm{Ch}}$) denotes the number of eigenstates in orbital $A$ ($B$) whose eigenvalues are smaller than $\alpha_0$.
Namely, for $\alpha<0$ ($\alpha>0$), $N_{\mathrm{ps0Ch}}$ of $h_{\mathrm{H}}\oplus rh_{\mathrm{H}}$ takes $1$ ($-1$).

Now, we consider the following correlated model 
\begin{subequations}
\begin{eqnarray}
\label{eq: Hcorr app}
\hat{H}_{\mathrm{H}}&=& \hat{H}_{\mathrm{H}0}+\hat{H}_{\mathrm{H}\mu}+\hat{H}_{\mathrm{H}V}, 
\end{eqnarray}
with
\begin{eqnarray}
\hat{H}_{\mathrm{H}\mu}&=& -\mu \hat{N}_{\uparrow} -r \mu \hat{N}_{\downarrow}, \\
\hat{H}_{\mathrm{H}V}&=& V (\hat{c}^\dagger_{a\uparrow}\hat{c}^\dagger_{a\downarrow}\hat{c}_{b\downarrow}\hat{c}_{b\uparrow}+h.c).
\end{eqnarray}
\end{subequations}
Here, $\mu$ and $V$ are real numbers.
We note that $\hat{H}_{\mathrm{H}\mu}$ and $\hat{H}_{\mathrm{H}V}$ are also invariant under applying $\hat{U}_{\mathrm{H}}$ which can be confirmed by recalling Eq.~(\ref{eq: U psi Udag app}).

\begin{figure}[!h]
\begin{minipage}{0.7\hsize}
\begin{center}
\includegraphics[width=1\hsize,clip]{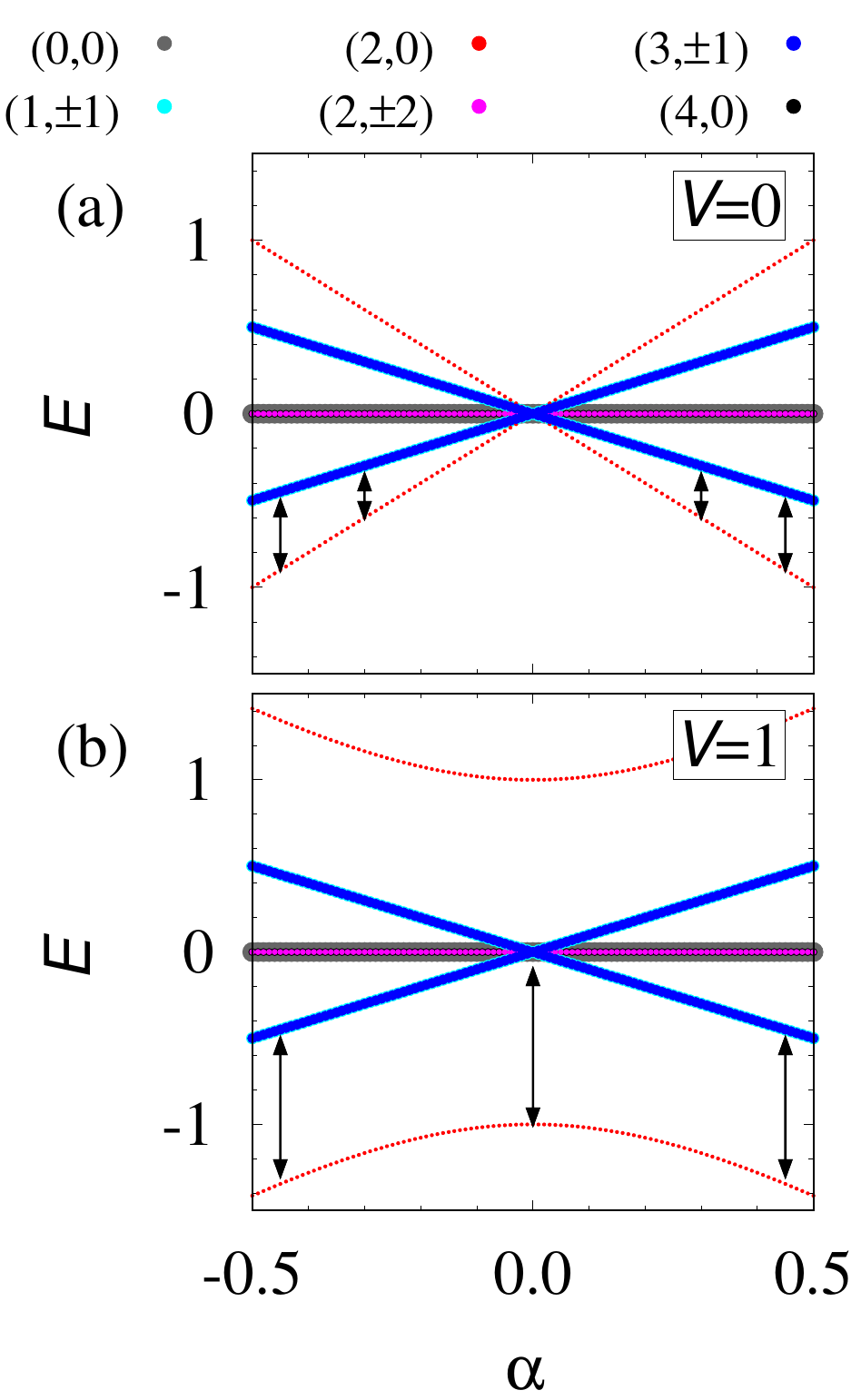}
\end{center}
\end{minipage}
\caption{
(a) [(b)]: Energy spectrum of the Hamiltonian $\hat{H}_{\mathrm{H}}$ as functions of $\alpha$ for $V=0$ [$V=1$].
Colored dot denote eigenvalues for subsectors with $(\hat{N}_{\mathrm{tot}},2\hat{S}^z_{\mathrm{tot}})$.
The each eigenvalue of subsector with $(\hat{N}_{\mathrm{tot}}, 2\hat{S}^z_{\mathrm{tot}})=(1,\pm1)$ is the same as that of subsector with $(\hat{N}_{\mathrm{tot}}, 2\hat{S}^z_{\mathrm{tot}})=(3,\pm1)$.
The arrows denote the gap.
The data are obtained for $r=0.5$ and $\mu=\alpha_0$.
}
\label{fig: Hermi}
\end{figure}

In the presence of correlation, the topological phase characterized by $N_{\mathrm{ps0Ch}}=1$ can be adiabatically connected to the phase characterized by $N_{\mathrm{ps0Ch}}=-1$.
To see this, let us start with the non-interacting case.
For $V=0$, the gap closes due to the topological phase transition described by the one-body Hamiltonian $h_{\mathrm{H}}$ [see Fig.~\ref{fig: Hermi}(a)].
In contrast to the non-interacting case, the system does not show the gap-closing for $V=1$ [see Fig.~\ref{fig: Hermi}(b)]. We note that each eigenstate preserves the symmetry described by $\hat{U}_{\mathrm{H}}$.

The above results indicate that the topological phase characterized by $N_{\mathrm{ps0Ch}}=1$ can be adiabatically connected to the phase characterized by $N_{\mathrm{ps0Ch}}=-1$ in the presence of the interaction.
In other words, the above two phases are topologically equivalent in the presence of correlations, which implies the reduction of topological classification $\mathbb{Z}\to \mathbb{Z}_{2}$.

We finish this part with a remark on the non-Hermitian case.
For non-Hermitian systems showing the non-trivial point-gap topology, we have defined the point-gap of the many-body Hamiltonian by introducing reference energy $E_{\mathrm{ref}}$, while for the Hermitian case, the gap is defined as the energy difference between the ground state and the first excited state.

\section{
Details of spectral flows for $\mu=0$
}
\label{sec: all spec mu 0 app}

\begin{figure}[!h]
\begin{minipage}{1\hsize}
\begin{center}
\includegraphics[width=1\hsize,clip]{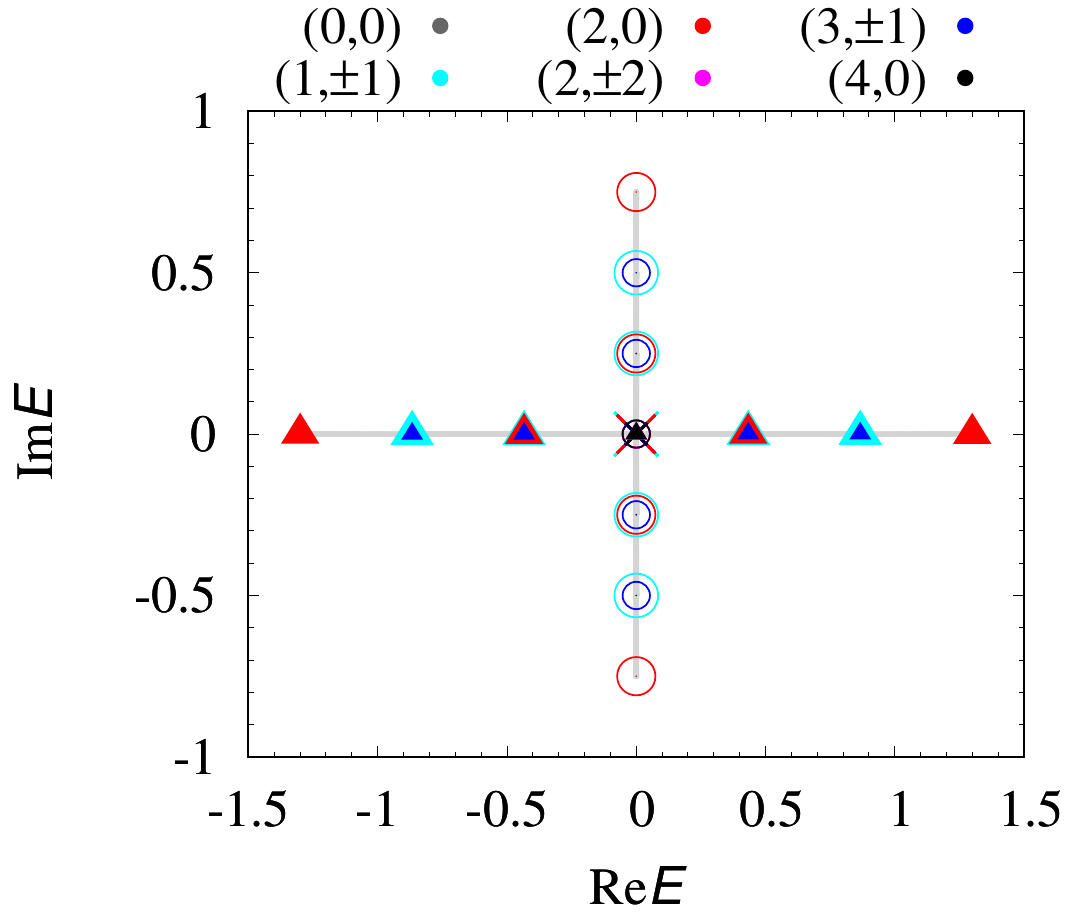}
\end{center}
\end{minipage}
\caption{
Spectral flow for each subsector with $(\hat{N}_{\mathrm{tot}},2\hat{S}^z_{\mathrm{tot}})$ at $\mu=0$ and $U=0$.
These data are plotted in a similar way as Fig.~\ref{fig: Eall U0 app}.
Circles, crosses, and triangles denote eigenvalues for $\beta=0$, $0.5$, and $1$, respectively.
These data are obtained for $\alpha=0.5$ and $r=0.5$.
}
\label{fig: Eall U0mu0 app}
\end{figure}

We discuss the spectral flow for $\mu=0$.
In Fig.~\ref{fig: Eall U0mu0 app}, the spectral flow for $U=0$ is plotted. As shown in this figure, subsectors labeled by $\hat{N}_{\mathrm{tot}}=1$, $\hat{N}_{\mathrm{tot}}=2$, and $\hat{N}_{\mathrm{tot}}=3$ are involved with the EP at $E=0$, which can be understood by analyzing the one-body Hamiltonian $h$.

\begin{figure}[!h]
\begin{minipage}{1\hsize}
\begin{center}
\includegraphics[width=1\hsize,clip]{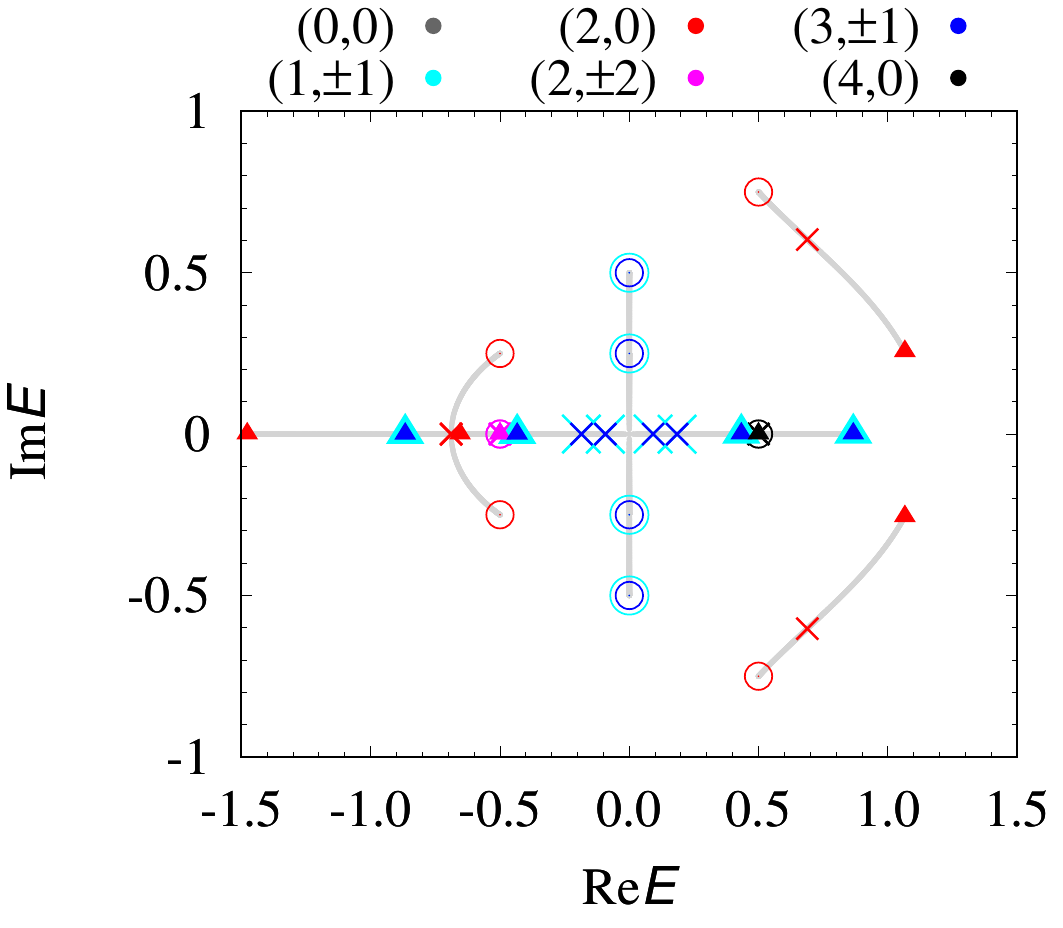}
\end{center}
\end{minipage}
\caption{
Spectral flow for each subsector with $(\hat{N}_{\mathrm{tot}},2\hat{S}^z_{\mathrm{tot}})$ at $\mu=0$ and $U=1$.
These data are plotted in a similar way as Fig.~\ref{fig: Eall U1 app}.
Circles, crosses, and triangles denote eigenvalues for $\beta=0$, $0.533$, and $1$, respectively.
These data are obtained for $\alpha=0.5$ and $r=0.5$.
}
\label{fig: Eall U1mu0 app}
\end{figure}

On the other hand, in the presence of the interaction, subsectors labeled by $\hat{N}_{\mathrm{tot}}=1$ and $\hat{N}_{\mathrm{tot}}=3$ are not involved with the MEP emerging away from the imaginary axis [see Fig.~\ref{fig: Eall U1mu0 app}].
This fact also supports that the MEP is described by spin degrees of freedom and essentially differs from the EP in the non-interacting case.

\end{document}